\documentclass[%
 aip,
 amsmath,amssymb,
 reprint,%
]{revtex4-2}

\usepackage{graphicx}
\usepackage{dcolumn}
\usepackage{bm}

\usepackage{color}
\usepackage[utf8]{inputenc}
\usepackage[T1]{fontenc}
\usepackage{mathptmx}
\usepackage{etoolbox}

\newcommand{\red}[1]{{#1}}
\newcommand{\blue}[1]{{#1}}

\makeatletter
\def\@email#1#2{%
 \endgroup
 \patchcmd{\titleblock@produce}
  {\frontmatter@RRAPformat}
  {\frontmatter@RRAPformat{\produce@RRAP{*#1\href{mailto:#2}{#2}}}\frontmatter@RRAPformat}
  {}{}
}%
\makeatother
\begin{document}

\preprint{AIP/123-QED}

\title[Delta-alpha cross-frequency coupling for different brain regions ]{Delta-alpha cross-frequency coupling for different brain regions}
\author{Dushko Lukarski}
\affiliation{Faculty of Medicine, Ss. Cyril and Methodius University, 1000 Skopje, Macedonia.}
 \affiliation{University Clinic for Radiotherapy and Oncology, 1000 Skopje, Macedonia.}

\author{Spase Petkoski}%
\affiliation{Aix Marseille Univ, INSERM, Inst Neurosci Syst (INS), 13005 Marseille, France}

\author{Peng Ji}
\affiliation{Institute of Science and Technology for Brain-Inspired Intelligence, Fudan University, 200433 Shanghai, China}

\author{Tomislav Stankovski}
 \affiliation{Faculty of Medicine, Ss. Cyril and Methodius University, 1000 Skopje, Macedonia.}
 \affiliation{Department of Physics, Lancaster University, LA1 4YB Lancaster, United Kingdom.}
 \email{t.stankovski@ukim.edu.mk.}

\date{\today}

\begin{abstract}
Neural interactions occur on different levels and scales. It is of particular importance to understand how  they are distributed among  different  neuroanatomical and physiological relevant brain regions. We investigated neural cross-frequency couplings between different brain regions according to the Desikan-Killiany brain parcellation. The adaptive dynamic Bayesian inference method was applied to EEG measurements of healthy resting subjects in order to reconstruct the coupling functions. \red{It was found that even after averaging over all subjects, the mean coupling function showed a characteristic waveform, confirming the direct influence of the delta-phase on the alpha-phase dynamics in certain brain regions and that the shape of the coupling function changes for different regions.} While the averaged coupling function within a region was of similar form, the region-averaged coupling function was averaged out, which implies that there is a common  dependence within separate regions across the subjects. It was also found that for certain regions the influence of delta on alpha oscillations is more pronounced and that \red{oscillations that influence other} are more evenly distributed across brain regions than the influenced oscillations. When \red{presenting} the information on brain lobes, it was shown that the influence of delta emanating from the brain as a whole is greatest on the alpha oscillations of the cingulate frontal lobe, and at the same time the influence of delta from the cingulate parietal brain lobe is greatest on the alpha oscillations of the whole brain.
\end{abstract}

\maketitle

\begin{quotation}
The delta-alpha cross-frequency coupling is proving to be a valuable descriptor in increasingly more brain states and domains. Here, by applying the adaptive dynamic Bayesian inference to EEG signals of subjects at rest we reconstructed the neural cross-frequency delta to alpha coupling functions that describe the interaction mechanisms of different regions of the brain. With this analysis framework we found a number of significant brain connections, as well as several characteristic differences between the brain regions.
\end{quotation}

\section{\label{sec:level1}Introduction}

The interactions in the brain are fundamental to the human ability to perceive and interact with the world. The brain is a heavily connected dynamical network system \cite{Park:13}, with interactions that are very complex and involve a vast network of neurons and synapses. Such complex system can mediate a vast number of functions, from a relatively static structure. Importantly, the brain can evolve with time, and different changes and transitions can occur \cite{Lehnertz:14,Chai:17}. Because not all the neurons and network processes in the brain are active at all time, and because they can exhibit collective, clustered, and synchronized behaviour \cite{Rudrauf:06,Li:22,Sauseng:08}, different types of changes, disruptions and transitions in the neural activity can occur \cite{Suo:15,Olde:14}.

Since the functions of the brain are highly dependent on its structure, and different functions are probably performed by different brain regions with different architecture, it is essential to identify the different regions of the brain in order to better understand its functions. For that reason, a significant effort has been invested by the scientific community in the direction of parcellation of the brain, starting from the classic Brodmann map, through the widely used Desikan-Killiany atlas \cite{Desikan:06}, all the way to the recently published human Brainnetome atlas \cite{Fan:16} and Human Connectome Project (HCP) multimodal parcellation \cite{Glasser:16} using \textit{in vivo} MRI data.

The brain connectivity is crucial to understand how the neurons and the brain dynamics evolve. A particularly accessible and useful approach has been the study of neural cross-frequency coupling, usually extracted from an electroencephalograph (EEG) recording \cite{Canolty:06,Jensen:07,Palva:05,Jirsa:13,Sorrentino:22}. Neural cross-frequency coupling refers to the interaction between different frequencies of neural brainwave oscillations in the brain. Cross-frequency coupling occurs when the amplitude or phase of one frequency band of oscillations is modulated by another frequency band. Thus, there are different types of cross-frequency coupling, such as amplitude-amplitude coupling, phase-phase coupling, and amplitude-phase coupling.

Neural cross-frequency coupling can be studied between different combinations of brainwave oscillations. In this work, we will focus on the delta-to-alpha neural cross-frequency coupling. Namely, it is well known that delta and alpha brainwave oscillations play an important role in the brain dynamics \cite{Delimayanti:20,Gorgoni:20,Krueger:16,Bashan:12,Penzel07b,Palva:07}. For instance, there are differences in frequency and power during different sleep stages which appear in the separate delta and alpha brainwave dynamics \cite{Ehlers:89,Keshavan:98,Benca:99,Scheuler:83} and in their related delta-alpha effect \cite{Hauri:73,Vijayan:15}. In another example, in a previous study about general anesthesia \cite{Stankovski:16} it was found that the delta-alpha coupling function is statistically significant and strong during anaesthesia. Similarly, previous works observe a prominent delta-alpha coupling in resting state \cite{Jirsa:13,Stankovski:17c}, during the orienting response \cite{Isler:08} and during sleep within the network physiology approach \cite{Bashan:12}. \red{A characteristic form of the delta-alpha coupling functions was also established \cite{Stankovski:16,Stankovski:17c, Manasova:23}.} These works point out that the choice to investigate the delta-to-alpha coupling among the brain regions had a relevance for the present study of resting state.

To perform the analysis needed we used comprehensive set of methods. First, to observe the oscillatory content of the brainwave oscillations we used wavelet time-frequency analysis. Then, we used the fact that the delta and alpha brainwaves have pronounced oscillating dynamics in order to study the interactions through their reduced phase dynamics \cite{Kuramoto:84}, thus observing phase-phase cross-frequency coupling. Here, we applied a method based on adaptive dynamical Bayesian inference for analysis of data to reconstruct a dynamical phase model describing the systems and their interactions  \cite{Smelyanskiy:05a,Stankovski:12b,Lukarski:20}.
The method of dynamical inference reconstructs effective connectivity \cite{Friston:11,Park:13} and reveals the underlying dynamical mechanisms. Here, we reconstruct the phase \emph{coupling functions} which describe how the interaction occurs and manifest, thus revealing a functional mechanism \cite{Stankovski:17b}. The design of powerful methods and the explicit assessment of coupling functions have led to applications in different scientific fields including chemistry \cite{Kiss:07}, climate \cite{Moon:19}, secure communications \cite{Stankovski:14a}, mechanics \cite{Kralemann:08}, social sciences \cite{Ranganathan:14}, and oscillatory interaction in physiology for cardiorespiratory interactions \cite{Kralemann:13b,Lukarski:22}. Arguably, the greatest recent interest for coupling functions is coming from neuroscience \cite{Stankovski:21}, where works have encompassed the theory and inference of a diversity of neural phenomena, physical regions, and physiological conditions \cite{Bick:20,Yeldesbay:19,Suzuki:18,Jafarian:19,Su:18,Stankovski:17c,Takembo:18,Gruszecka:21}.

\blue{\section{\label{sec:level1}Materials and Methods}}

\subsection{\label{sec:level2}Adaptive Dynamical Bayesian Inference}

When investigating a complex dynamical oscillatory system, such as the oscillatory behaviour of the brain, one way to gain new insights is by modeling the system by using differential equations. Usually, by measuring some signals originating from the oscillatory time evolution of the system, one can infer the parameters of a model that describes the system under certain conditions. According to the phase reduction theory, in case when the interactions between the oscillators are sufficiently weak, the behaviour of the system can be approximated with its phase dynamics \cite{Kuramoto:84,Nakao:14,Rodrigues:16}. If the phases of the system can be considered as monotonic change of the variables, \red{the partial dynamical process of the node $i$ as influenced by another node $j$} can be represented with the system of differential equations:
\begin{equation}
\dot{\varphi}_{i,j}=\omega_i+q_{i,j}(\varphi_i,\varphi_j)+\xi_i,\label{eq:01}
\end{equation}
where $\varphi_i$ is the phase of the i-th oscillator, $\omega_i$ is its angular frequency parameter, $q_{i,j}$ is the coupling function which describes the influence of  the j-th oscillator on the i-th oscillator, and $\xi_i$  represents the noise. Usually, the noise is assumed as a white Gaussian noise given by $\xi_i(t)\xi_j(\tau)=\delta(t-\tau)E_{ij}$, where the information about the correlation between the noises of the different oscillators is included in the symmetric matrix $E_{ij}$. \red{In theory, the full model for the phase dynamics of a brain region oscillator should contain all the connections at once, in a single phase equation. However, due to the high dimensionality and computational expense, with equation (1) we infer a partial part of the full model dynamics related only to the two brain regions involved in a coupling connection. This procedure is then applied for each pair of brain regions.}

Because of the periodic nature of the system, the coupling function can be represented by a Fourier decomposition:
\begin{equation}
q_{i,j}(\varphi_i,\varphi_j)=\sum_{k=-\infty}^{\infty}\sum_{s=-\infty}^{\infty}c_{i;k,s}e^{i2{\pi}k \varphi_i}e^{i2{\pi}s\varphi_j}. \label{eq:02}
\end{equation}
For a system of two coupled oscillators, reduction to a finite number K of Fourier terms will give:
\begin{equation}
\dot{\varphi}_{i,j}=\sum_{k=-K}^{K}{c_k}^i\Phi_{i,j,k}(\varphi_i,\varphi_j)+\xi_i(t),\label{eq:03}
\end{equation}
where$ {\Phi_{i,j,0}=1}, {{c_0}^i=\omega_i}$, and the rest ${\Phi_{i,j,k}}$ and ${{c_k}^i}$ are the $K$ most important Fourier components (in this work we use $K=2$).
With the assumption for a white Gaussian noise, the task is then reduced to inference of the unknown parameters of the model:
\begin{equation}
M=\left\{{c_k}^i,E_{ij}\right\}.\label{eq:04}
\end{equation}

When the parameters of the model are inferred, one can  determine then the coupling functions $q_{i,j}$ which describe the underlying mechanisms of the interaction of the oscillators \cite{Stankovski:17b}.

In this work we employ the method of adaptive dynamical Bayesian inference (aDBI) \cite{Stankovski:12b,Duggento:12,Stankovski:14d, Lukarski:20} in order to gain new insights into the oscillatory behaviour of the brain regions and the brain lobes. In this method, the time series of phases of the oscillators are considered to be time sequences of blocks of samples. In each block, the samples from a certain time interval are included, and the duration of this time interval is specified by the time window $t_w$. In the inference procedure, the initial assumptions for the parameters of the model are that ${c_k}^i=0$, and therefore at least few inference blocks are required to obtain appropriate estimates of the model parameters values and the corresponding coupling functions. To obtain improved inference in every subsequent block, part of the information from the previous block is included in the following one. The so-called propagation parameter $p_w$ controls how much of the information of the previous block is included in the following one. In the method of aDBI both the time window $t_w$ and the propagation parameter $p_w$ are adaptively determined, based on the time variabilities present in the signal. After determination of $t_w$ and $p_w$ the final inference is performed. This final inference provides the values of the parameters and the coupling functions for each block of inference, thus observing the time evolution of the system and the inter interactions, with a temporal resolution defined by the time window $t_w$. \red{Further technical details about the parametrization, convergence and robustness of the aDBI method can be found in previous papers \cite{Duggento:12,Stankovski:14d,Lukarski:20,Stankovski:12b}.} Even though the aDBI was introduced for studies of coupled phase oscillators with oscillating frequencies in the cardiorespiratory range, the procedures is applicable to the frequency range of the brain waves as well. The application of the aDBI method on the subject dataset that was used in this study yielded a time window of $t_w=10$ s and a propagation parameter of $p_w=0.2$.

\red{The aDBI method represents a further improvement of the DBI method \cite{Stankovski:12b,Duggento:12}, by minimizing the covariance matrix which is an indicator of the quality of the inference. The details of the method are given elsewhere \cite{Lukarski:20} and in summary it leads to an improved inference of the model parameters without losing information about the temporal changes in the behavior of the oscillators.}

\begin{figure}
\includegraphics[width=3.1in]{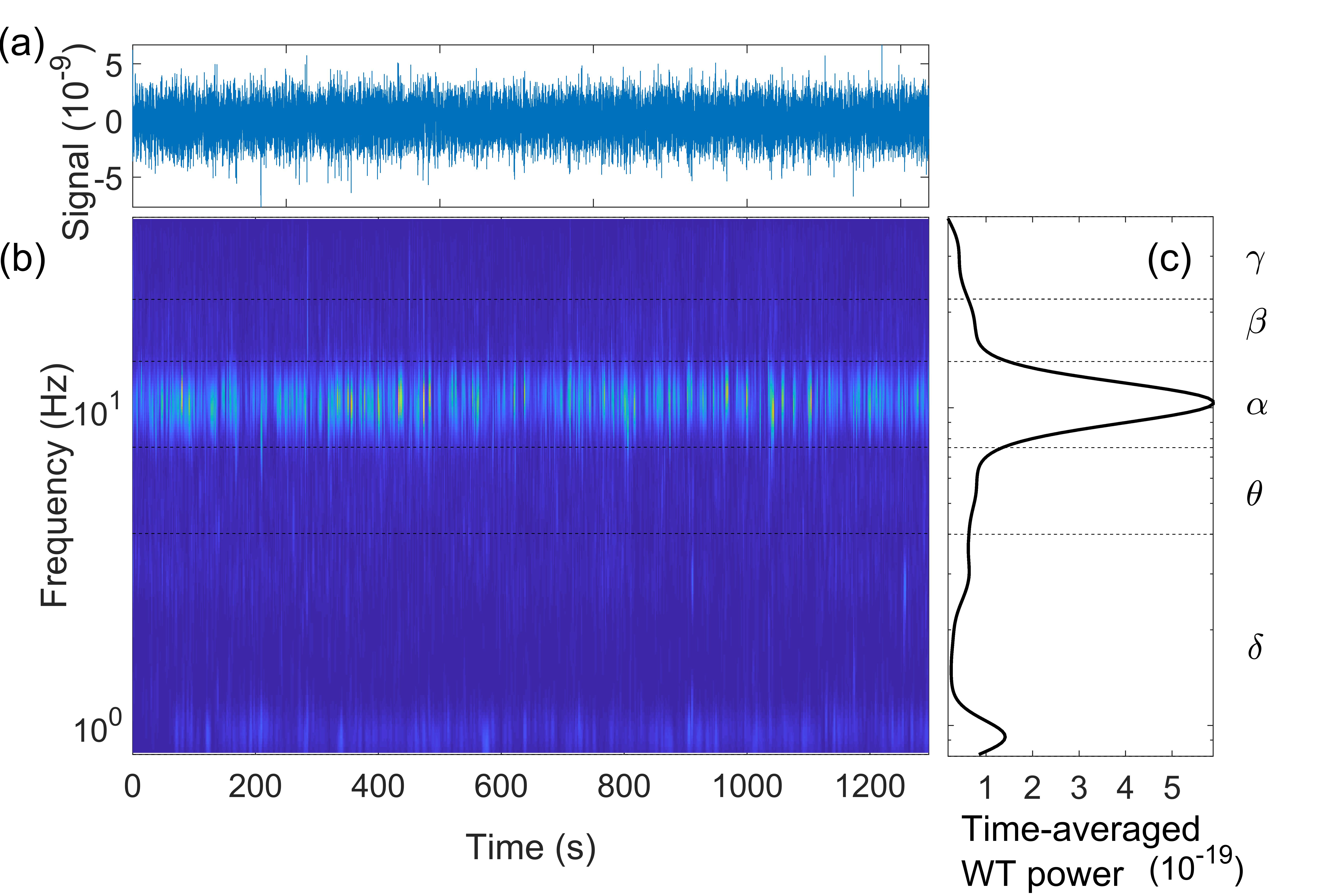}
\caption{\label{Fig_1_WT} Wavelet transform of the measured signal for one of the regions of one subject. The measured signal is shown in (a), while the corresponding time-frequency wavelet transform is shown in (b). The time-averaged intensity of the wavelet is shown in (c).}
\end{figure}

\begin{figure*}
\includegraphics[width=\textwidth]{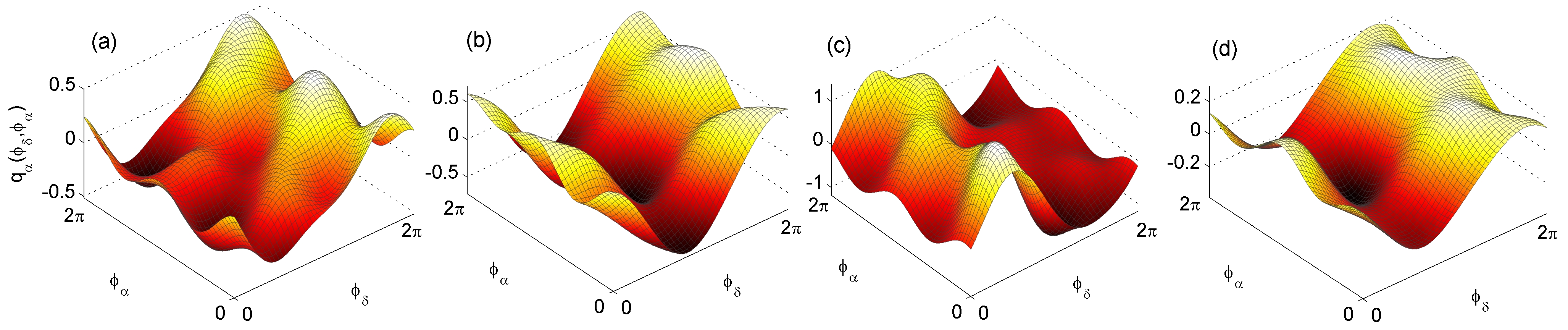}
\caption{\label{Fig_2_CF}The delta-alpha neural coupling functions. Examples of individual subject delta-alpha coupling functions between different regions (a)-(c). In particular, (a) shows a coupling function from subject 10 between delta region 60 and alpha region 60, (b) from subject 12 between delta region 50 and alpha region 23, and (c) from subject 6 between delta region 54 and alpha region 19. The last plot (d) shows a subject-averaged delta-alpha coupling function between delta region 58 and alpha region 21.}
\end{figure*}

\blue{\subsection{\label{sec:level2}Dataset}}

The dataset used in this study is publicly available \cite{Schirner:18} (https://osf.io/mndt8/). The data source contains the empirical region-average fMRI (functional magnetic resonance imaging), EEG source activity and structural connectomes of the $68$ parcellated cortical regions of the brain of 15 healthy human subjects, age 18-31, eight of whom are female.\red{The data consists of a resting state time series, where the subjects were asked to just stay awake and keep their eyes closed.} In this study we used the empirical EEG source activity and the structural connectomes. The time series of the source activity for each patient and each cortical region have duration of $21.6$ minutes with sampling frequency of $200$ Hz. The description of the $68$ parcellated cortical regions is given in the Appendix.\\

\blue{
\subsection{\label{sec:level2}Wavelet transform and the phase extraction procedure}
}

In order to check the existence of brain wave oscillations \red{and their frequency content}, the EEG time-series signals were first analyzed using a continuous wavelet transform \cite{Daubechies:92,Kaiser:94,Stefanovska:99b}. The continuous wavelet transform is defined by the equation
\begin{equation}
 WT(\omega,t)=\int_{0}^{\infty} \psi(\omega(u-t))x(u)\omega du,
\label{eq:6}
\end{equation}
where $x(t)$ is the signal, $\omega$ denotes the angular frequency, $t$ is the time and $$\psi(u) = \frac{1}{2\pi} (e^{i2\pi f_0 u} - e^{\frac{(2\pi f_0) ^2}{2}})e^{-\frac{u^2}{2}}$$ is the complex Morlet wavelet, with central frequency $f_0 = 1$, $\int \psi(t)dt = 0$, and with $i$ being the imaginary unit. The continuous wavelet transform is a time-frequency representation which contains both the phase and the amplitude dynamics of the oscillatory elements from the analyzed signal.

The initial wavelet observation of the oscillations contained in the corresponding EEG signals was carried out for several brain regions. After the initial wavelet observations, a phase extraction procedure was performed for the delta and alpha waves of the EEG signal in order to obtain the instantaneous phase time-series. These phase time-series then act as an input to the aDBI method. The oscillatory intervals were first evaluated by standard digital filtering procedure including FIR filter followed by a zero phase filtering procedure (``filtfilt") to ensure that no time or phase lags are introduced with the filtering procedure. The delta waves signal limits were from 1 to 4 Hz, while the alpha waves signal limits were from 8 to 12 Hz \cite{Buzsaki:04}.
The phases of the filtered signals were estimated via Hilbert transformation, thus obtaining the protophases. On these protophases, the protophase-to-phase transformation was applied \cite{Kralemann:08} in order to obtain the independent phases \red{which act as input signals for the
Bayesian inference}.\\

\blue{\subsection{\label{sec:level2} Surrogate data testing}}

When oscillatory signals are analyzed, the inferred coupling between the signals is always positive and non-zero, even if the oscillators are uncoupled or unrelated. Therefore, it is necessary to establish a significance threshold in order to determine if the obtained coupling strength indicates a genuine connection and interdependence of the phenomena. \red{Such surrogate data is used for statistical testing of the coupling strength.} A threshold is usually defined by constructing randomized surrogates of the original signals \cite{Schreiber:00b,Lancaster:18a} and calculating the coupling strength for these surrogates. The coupling strength obtained in this manner represent a baseline for the confirmation of the coupling of the oscillators. 
In this work surrogates were constructed for each of the $68 \times 68$ delta-alpha couplings going from, and to, each of the $68$ regions of the brain by using a procedure called cyclic phase permutation surrogates \cite{Lancaster:18a}  based on rearrangement of the cycles within the extracted phase. The surrogate threshold taken in this work is the mean plus two standard deviations (mean + 2STD) of the surrogate couplings.\\

\section{\label{sec:level1}Results}

Fig.~\ref{Fig_1_WT} shows the wavelet transform of the measured signal for one of the $68$ brain regions in one of the subjects. The signal itself is presented in Fig.~\ref{Fig_1_WT} (a), while the corresponding time-frequency wavelet transform is shown in Fig.~\ref{Fig_1_WT} (b). To show the oscillatory frequencies present in the signal more clearly, the time-averaged intensity of the wavelet is presented in Fig.~\ref{Fig_1_WT} (c). The frequency intervals of the corresponding brain waves are given with the dashed lines. From the figure, one can clearly see the strong alpha wave, as well as the delta wave with a slightly lower wavelet power.

The delta-alpha coupling functions are presented in Fig.~\ref{Fig_2_CF}. The coupling functions are evaluated first on individual subjects for specific regions -- Fig.~\ref{Fig_2_CF} (a)-(c). Here, they show the characteristic functional form where the delta-alpha phase coupling function depends predominately on the delta dynamics, or in other words it reflects the direct influence that the delta phase dynamics exert on the alpha phase dynamics by accelerating or decelerating the alpha brainwave oscillations. \red{This specific form belongs to the category of direct, among the separation of self, direct and common coupling functions \cite{Stankovski:15a,Iatsenko:13a}}. By comparing the three plots for the coupling functions Fig.~\ref{Fig_2_CF} (a), (b) and (c), one can notice  that this direct influence is like a wave that shifts from left to right from $0$ to $2\pi$ along the delta axis. In general, it keeps the direct delta dependence (i.e.\ it still changes predominately on delta axis) but it shifts  the maximum of the function along the delta axis.

When we average the coupling functions for the same region from all the subjects, as shown in Fig.~\ref{Fig_2_CF} (d), the remaining average delta-alpha coupling function still reflects the direct delta dependence, albeit with slightly reduced amplitude due to averaging. \red{The Appendix A shows how this coupling function is similar or different in respect to all the order regions.} 
Furthermore, when we average all the coupling functions across regions and subjects, the average coupling function disappeared i.e.\ it was insignificantly low without a common form of the function. In other words the region average coupling function averaged out, because there was no specific common form between regions.

\begin{figure}
\includegraphics[width=3.4in]{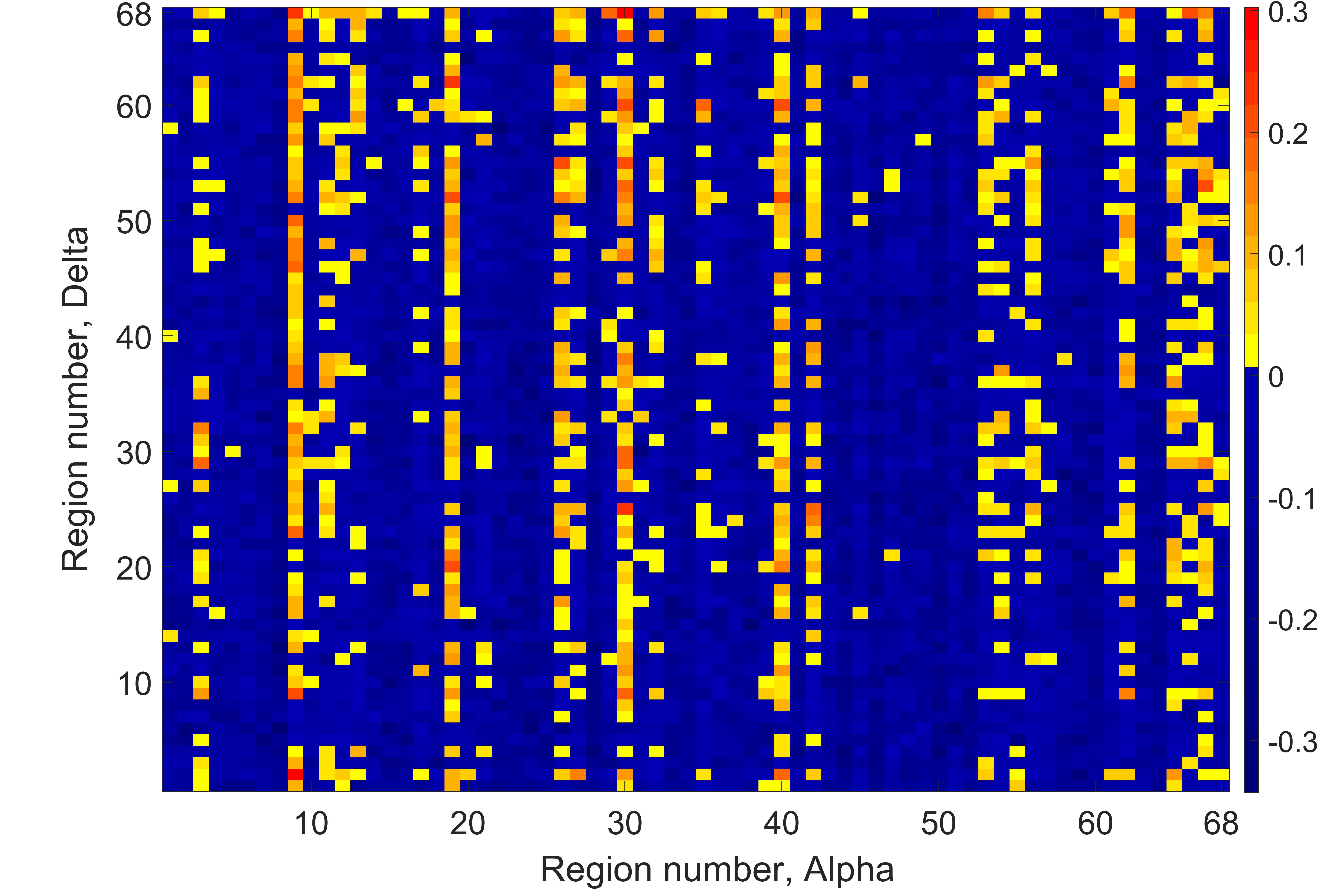}
\caption{\label{Fig_3_Matr} Matrix $68 \times 68$ showing the delta-alpha couplings from region to region. Only those couplings which are statistically significant in respect of the surrogate threshold are shown in color (values above $0$). }
\end{figure}

\begin{figure}
\includegraphics[width=3.4in]{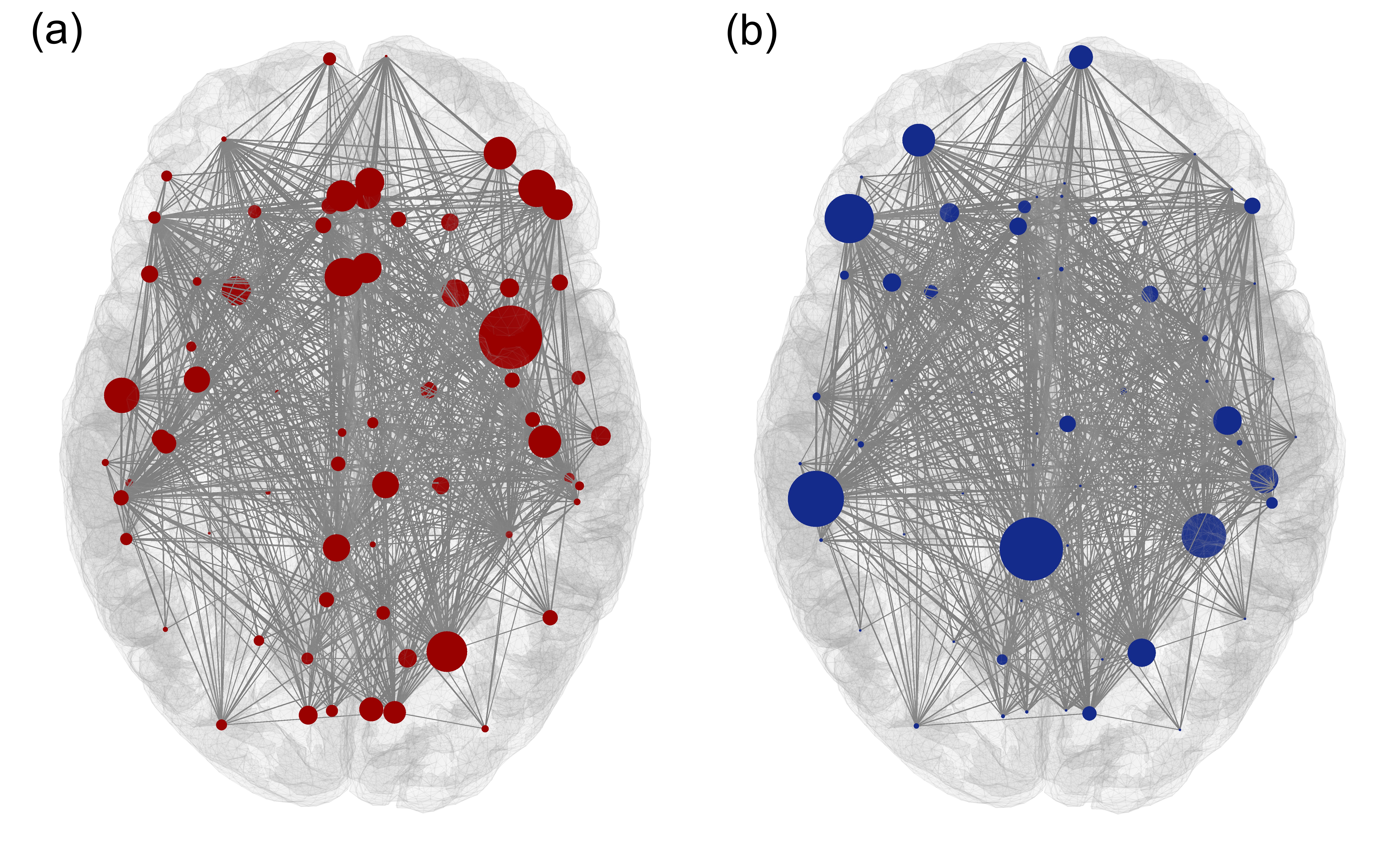}
\caption{\label{Fig_4_brain}Significant delta-alpha coupling strengths for the different brain regions. (a) The radii of the red circles correspond to the sum of coupling strengths of all the delta exiting the corresponding region. (b) The radii of the blue circles correspond to the sum of coupling strengths of all the alpha entering the corresponding region of the brain.}
\end{figure}

Fig.~\ref{Fig_3_Matr} shows a $68 \times 68$ matrix representing the significant delta-alpha coupling functions for all the $68$ brain regions. The vertical axis shows the number of the region for the delta brainwaves, while on the horizontal axis the number of the region of the alpha brainwaves is given. \red{The matrix is not symmetrical and the coupling indicated by the columns is different than the one in the rows. The figure indicates that for some regions of the brain (e.g.\ columns $9, 19, 30, 40$, etc) there is a stronger influence from the delta waves to the alpha waves.} This is also shown in Fig.~\ref{Fig_4_brain}, where these $68$ regions are marked as circles on a cross-section of the brain.	

Fig.~\ref{Fig_4_brain} (a) shows the summarized delta-alpha coupling strengths coming from a specific regions with red circles, while Fig.~\ref{Fig_4_brain} (b) shows the sum of the delta-alpha coupling strengths for the alpha of a specific regions with blue circles. The radii of the circles are proportional to the sum of the corresponding coupling strengths. One can notice that while the significant delta-alpha couplings emanate from various different brain regions, they end up in much smaller number of the brain regions.

\begin{figure}
\includegraphics[width=3.4in]{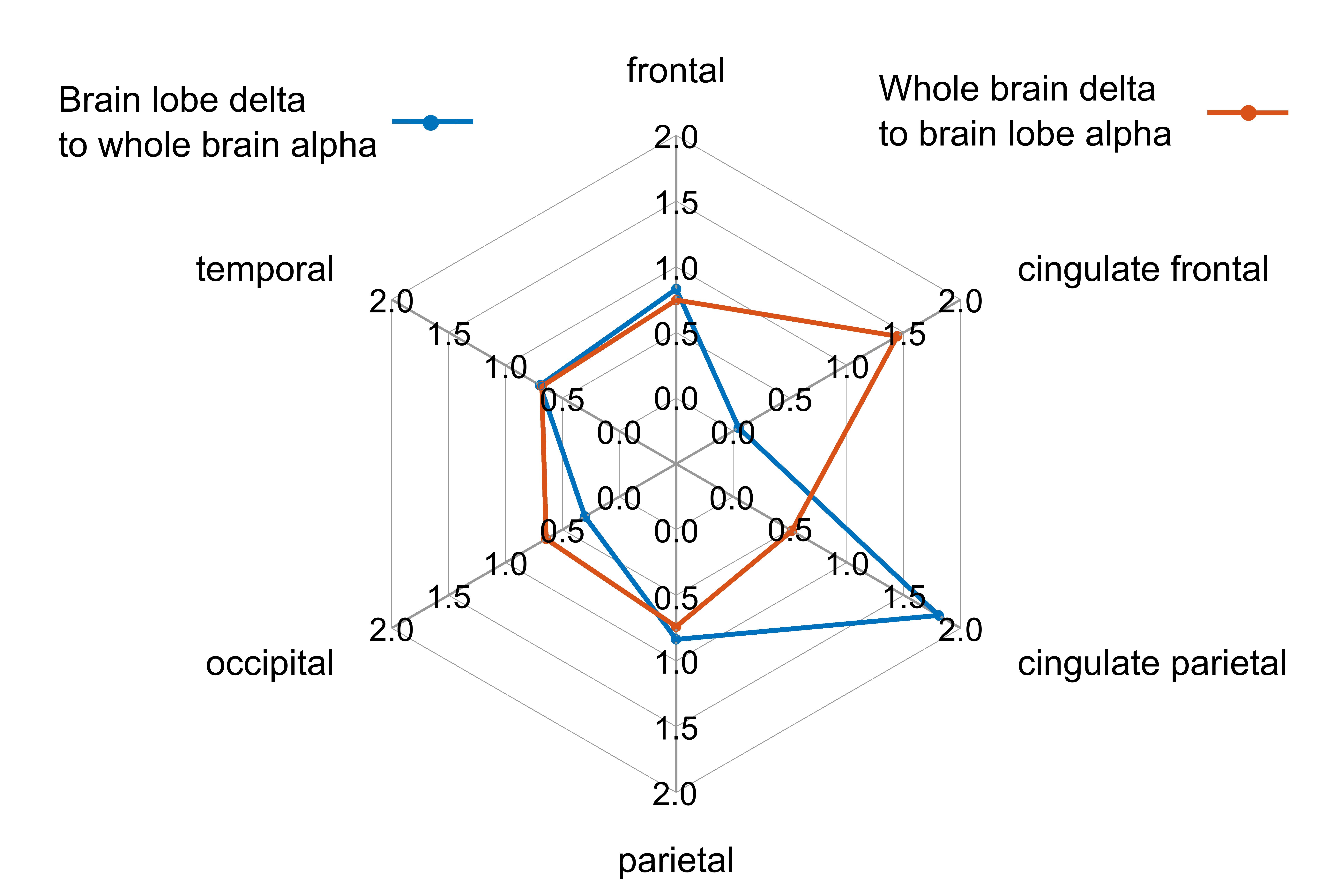}
\caption{\label{Fig_5_lobes} Spider plot showing the influence of the separate brainwaves of the interaction into the specific brain lobes. The influence of the delta from the entire brain to the alpha of specific brain lobes (blue curve) and the influence of delta of specific brain lobes to the alpha from the entire brain (red curve). }
\end{figure}

In order to obtain more tangible information about the overall interactions between the different brain lobes (frontal, cingulate frontal, cingulate parietal, parietal, occipital and temporal lobe), a summation of the significant coupling functions by brain lobes was performed. The sums obtained were normalized by the number of regions involved in each of the brain lobes. The results are presented in Fig.~\ref{Fig_5_lobes}. The normalized sum of the delta-alpha couplings, where the delta is from specific brain lobe and alpha from any lobe of the brain (whole brain alpha), is shown with blue line. While the normalized sum of the delta-alpha couplings, where the delta comes from any lobe of the brain (whole brain delta) and alpha from specific brain lobe, is shown with red line. From the spider plot (Fig.~\ref{Fig_5_lobes}) it can be seen that the greatest influence on the whole brain alpha has the cingulate parietal delta and at the same time the greatest influence of the whole brain delta is on the cingulate frontal alpha.

\section{\label{sec:level1} Discussion and Conclusions}

The influence of delta brain waves on alpha brain waves for a resting subject has previously been determined at the whole brain level. In this paper we try to gain a deeper insight by investigating this delta-alpha influence for different brain regions according to the Desikan-Kiliany anatomical parcellation of the brain \cite{Stankovski:17c,Jirsa:13,Isler:08,Bashan:12}. As presented in the results, it can be concluded that this influence is clearly visible for different regions, because even after averaging the delta-alpha coupling functions for a particular region across all the subjects, the mean coupling function still shows the characteristic shape (Fig. 2), confirming the direct influence of the delta-phase dynamics on the alpha-phase dynamics in certain brain regions. This influence consists in acceleration or deceleration of alpha oscillations under the influence of delta oscillations.

\red{In terms of analysis we have applied a comprehensive methodological framework for interacting brainwave oscillations. The nature of delta and alpha oscillations were observed with wavelet transform using standard parameters, with $f_0=1$ central frequency. This is a simple standard widely used procedure for time-frequency analysis.  For the reconstruction of the phase model we used the adaptive Bayesian inference. It is a well established method which has been widely used and tested for robustness and convergence, where its parametrization has been systematically investigated on different numerical and biological systems \cite{Duggento:12,Stankovski:14d,Lukarski:20,Stankovski:12b}. For verifying the statistical significance of the inferred delta-alpha coupling we have applied surrogate data testing \cite{Lancaster:18a}. 

The model  equation (\ref{eq:01}) assumes pairwise interaction between two regions and includes only coupling function with two phase variables. This is a simplified approximation, as the brain regions form parts of a complex network, and the full model of a phase oscillator should include all the brain connections in a single equation. With equation (\ref{eq:01}) we have thus separated the inference for a partial dynamics on two-by-two basis for all the pairs of brain regions. This was possible because the Bayesian method can allow such partial dynamical filtering. While the reason to do this and separate the inference was due to the high dimensionality (68x68 regions) and the computational complexity, which otherwise could have lead to problems such as parameters overfitting. Also, we have used only pairwise coupling functions. Thus, a natural extension of this work could include also non-pairwise multivariate coupling functions \cite{Battiston:20,Stankovski:15a}. This is a case where the coupling function in the dynamics of one region will have more than two phase variables from phases of other regions.
}

The coupling function results demonstrated that there is a common waveform, predominantly due to direct influence from delta oscillations, but the wave shifts along the delta axis for different regions -- Fig. 2 (a-c). We present three characteristic regions here, but the general observation from all the regions was that the wave was shifting for different regions. \red{The quantitative analysis in Appendix A also supported this by showing relative variations of the form for different regions}. The answer to why the wave for coupling function forms shifts for different regions might  be because there are different lengths for the structural pathways, through which  different regions interact. This  most likely implies different time delays for the signal propagation \cite{Sorrentino2021, lemarechal2021brain}, which is known to impact the synchronization and phase arrangement between brain regions \cite{Petkoski2019} and is crucial for the information transfer \cite{Fries2015}. This time delay manifests itself as a phase shift for the oscillatory activity, i.e.\ $\Delta\phi$ within the phase coupling functions (e.g. as in $q_\alpha(\phi_\delta,\phi_\alpha)=\epsilon \sin(\phi_\alpha-\phi_\delta-\Delta\phi)$), which in turn can be the cause of the phase shift of the wave observed in the figures. Our current initial observation of the structural and time delay information in this direction can stimulate future systematic analysis for quantifying how the space-time structure of the brain regions, as defined by the weights and time-delays of the connectome \cite{Petkoski2022}, impacts the resulting coupling functions. Such a question is even more valid because the averaged coupling function within a region was of similar form Fig. 2 (d), while the region-averaged coupling function was averaged out. The latter implies that there is a common deterministic dependence within regions across the subjects, which is different for the separate brain regions. This kind of analysis would first require better identification of the propagation velocities and the time-delays on a personalized level, which is still not established beside promising results of MRI as a myelin biomarker \cite{mancini2020interactive}, and proposed \textit{in vivo} techniques \cite{Sorrentino2021}. However, our results indicate that even aggregated atlases for time-delays \cite{lemarechal2021brain} could be useful, since some of the patterns are consistent across the subjects.

We have seen that this influence is not evenly distributed across brain regions, but for certain brain regions the influence of delta oscillations on alpha oscillations is more pronounced, as is the case for isthmus cingulate, pars triangularis (associated with verbal and non-verbal communication \cite{Foundas:96, Mai:11, Johns:14}) and the supramarginal region of the left hemisphere (involved in language processing \cite{Deschamps:14, Leinweber:14} and tool use action \cite{Ma:11, Assmus:03}) and the fusiform region of the right hemisphere (involved in object and face recognition \cite{Zhao:18, Katanoda:16, Axelrod:15}). To lesser extent this is also noticeable for the lateral orbitofrontal and the rostral middle frontal region of the left hemisphere and the inferior temporal, pars triangularis, posterior cingulate, superior parietal, frontal pole, temporal pole and transverse temporal region of the right hemisphere (see Fig.3). These regions are located in different lobes of the brain, most of them in the frontal and temporal lobes, but some in the parietal and cingulate parietal as well. No clear distinctions can be made in terms of the brain hemisphere, as is expected, since the different brain centers responsible for different actions are located in the different hemispheres.

This uneven distribution of delta influence on alpha oscillations from different regions is less pronounced on the delta side of the regions, as shown in Fig. 3 and more specifically in Fig. 4 (a). Fig. 4 shows that while the delta-alpha influence is more concentrated for the alpha waves in certain regions, the distributions of significant couplings in terms of delta waves is more even across brain regions. This means that the influencing oscillations are more evenly distributed across the brain regions then the influenced oscillations, which are more concentrated in certain regions.

Additional insights into delta-alpha influences in the brain can be gained by condensing this information down to the level of brain lobes, as shown in the spider plot  (Fig. 5). These results indicate that the influence of delta oscillations of all brain regions is greatest on alpha oscillations of the cingulate frontal lobe. At the same time, the influence of cingulate parietal brain lobe’s delta oscillations on the alpha oscillations is greatest among all the regions of the brain. This influence of the cingulate frontal and cingulate parietal regions of the brain on other brain regions and on the brain as a whole should be further investigated and put into the context of the functioning of the brain from a neurological point of view.

Finally, it is worth noting that we presented the methodological framework for interactions in the brain regions network for the resting state, however, the framework carries important implications and can readily be used also for other neural states, or interacting oscillatory networks more generally.

\begin{acknowledgments}
D.L., T.S. and P.J.  acknowledge support from the bilateral Macedonian - Chinese project for scientific
and technological cooperation.
P.J. acknowledge support from STI2030-Major Projects (2021ZD0204500),  the NSFC (62076071), and Shanghai Municipal Science and Technology Major Project (2018SHZDZX01).

\end{acknowledgments}

\section*{Data Availability Statement}

The data that support the findings of this study are publicly available.

\section{REFERENCES}


%


\appendix
\red{
\section{Similarity index for the inferred coupling functions}

The analysis about the coupling functions in Fig.\ \ref{Fig_2_CF} shows qualitatively that there is a similar form for different regions for some subjects and for subject average on some regions. To quantify how big is this similarity and how much there is variations and differences from the coupling functions presented, now we calculate  the similarity in respect to all the other regions.

One way to quantify the form of the coupling functions is to use the so-called similarity index \cite{Kralemann:13b} between the coupling functions. The similarity index measure, $\rho$, gives the similarity between two coupling functions $q_{1}$ and $q_{2}$, regardless of their coupling strengths. Thus, the similarity index $\rho$ is a coupling function unique measure that quantifies the form of the functions. It is calculated as correlation between the coefficients from the inferred coupling parameters \cite{Kralemann:13b}. The index is determined as:
\begin{equation}
\rho=\frac{\langle {\tilde q}_{1}{\tilde q}_{2}\rangle }{\|{\tilde q}_{1}\|\|{\tilde q}_{2}\|}
\end{equation}
where $\langle q \rangle$ denotes spatial averaging over a two dimensional domain $0 \leq \varphi_{1}, \varphi_{2} \leq 2\pi$, and $\tilde q = q- \langle q \rangle$ and $\| q \| = \langle qq \rangle ^{1/2}$. 

\begin{figure}
\includegraphics[width=3.4in]{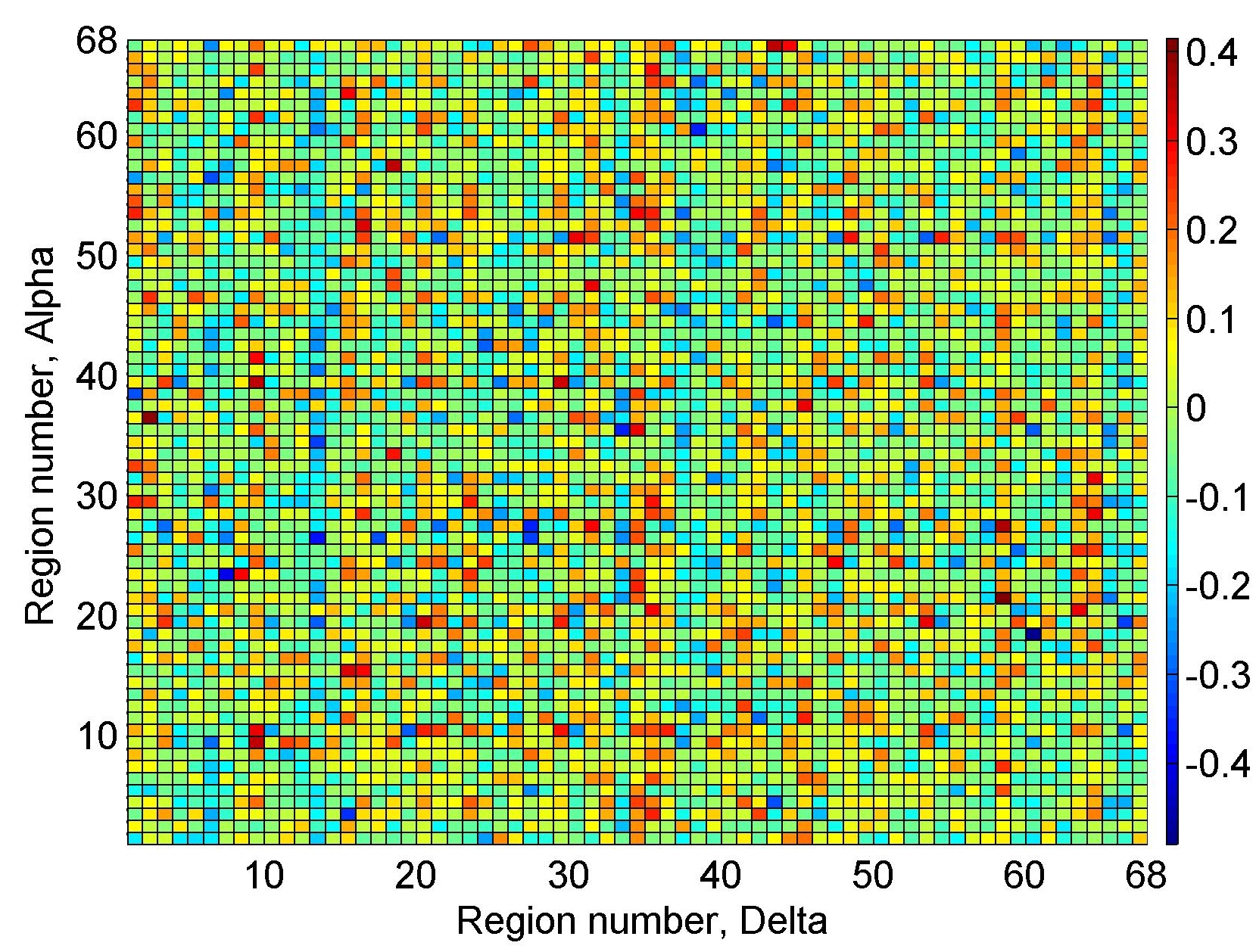}
\caption{\label{Fig_6_CFsim} Matrix $68 \times 68$ showing the similarity index for different regions of the brain. The similarity index shows the average similarity for the 15 subjects between the coupling function  for a particular brain region and the average coupling function for all subjects between delta region 58 and alpha region 21.}
\end{figure}

The results for the similarity index $\rho$ for the inferred coupling functions across regions are shown in Fig.~\ref{Fig_6_CFsim}. The  indices presented show the average similarity for all the subjects between the coupling functions of each region and the average coupling function for all subjects for a characteristic case between delta region 58 and alpha region 21 i.e.\ the coupling function as presented in Fig.\ \ref{Fig_2_CF} (d). Or in other words, because we could not present visually all the 68x68=4624 coupling function combinations, in the main text we present some characteristic cases, and here in the Appendix with Fig.\ \ref{Fig_6_CFsim} we extend this by quantitative analysis with all the other cases. Thus, Fig.\ \ref{Fig_6_CFsim} shows the extent of similarity and deviations from the visually presented coupling functions. From the observations, one can notice that there are relative variations of the form, with some regions being more or less similar, and there is also negative similarity i.e.\ $\pi$-shifted similarity.

\section{Sample size robustness -- variance of the coupling functions in respect to number of subjects}

\begin{figure}[h!]
\includegraphics[width=3.4in]{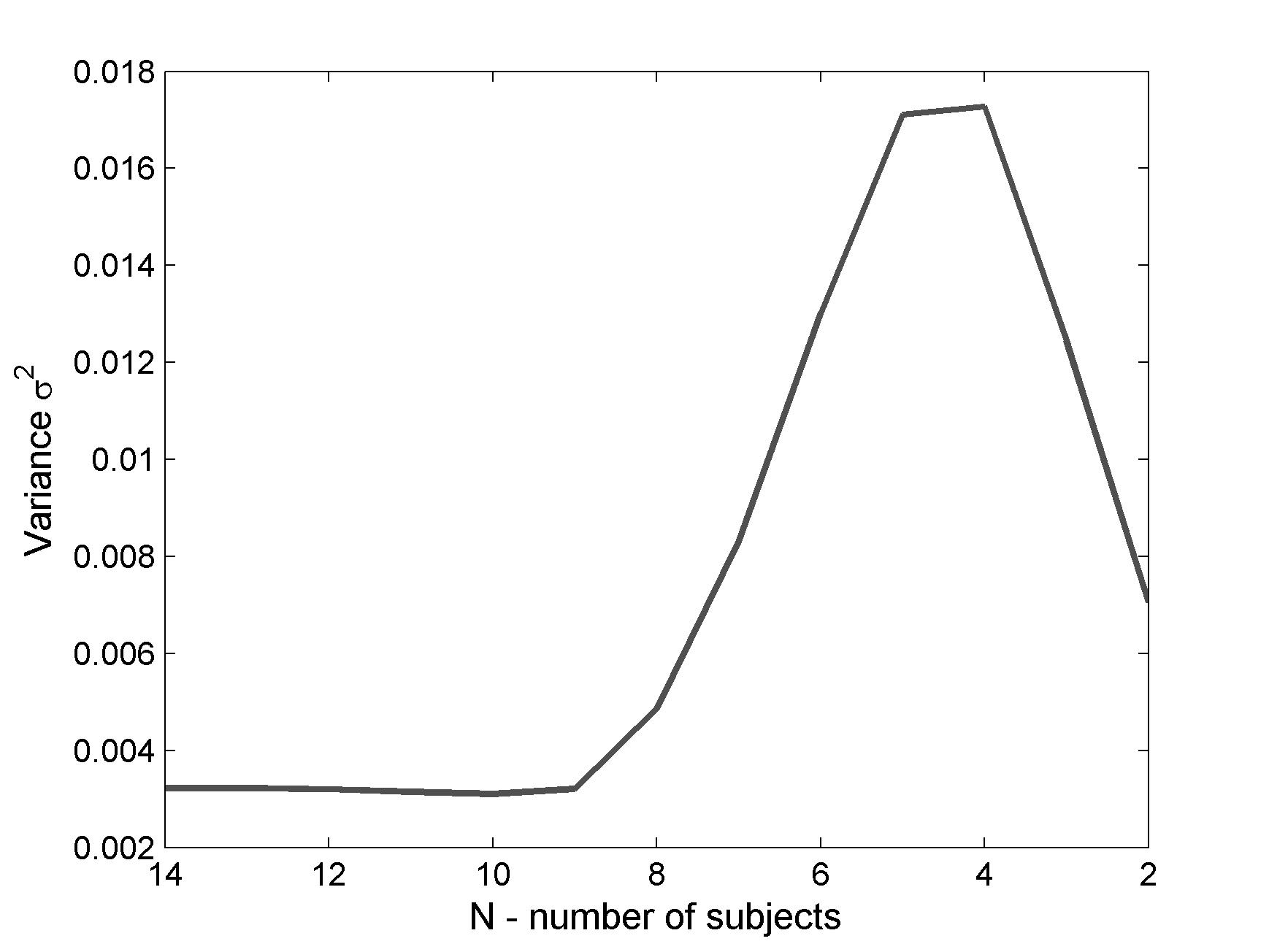}
\caption{\label{Fig_7var} Sample-size effect on the subject-averaged delta-alpha coupling function. The plot shows dependence of the variance in respect to the number N of subjects averaged. The variance is calculated from the similarity index between the full (N=15) subject averaged coupling function and the coupling functions from all the combinations of lower N number of subjects. All the coupling functions are for delta region 58 and alpha region 21.  }
\end{figure}

The coupling function analysis were perform on sample of N=15 subjects. In order to test if the number of subjects had an effect on the resulting average coupling function (like e.g.\ in Fig.\ \ref{Fig_2_CF}(d)) we tested how much the coupling function varies when smaller number of subjects is averaged. This was done by systematically calculating the average coupling function from smaller number of subjects N=14, then N=13, and so on until N=2. Here, for each smaller N we calculated the average coupling function for all N-combinations. Then, we calculated the similarity index $\rho$ between each N average coupling function and the coupling function from all 15 subjects (as in Fig.\ \ref{Fig_2_CF}(d)). Finally, we calculated the variance for all similarity indexes of each combination for one N. For example for N=14, there are 15 different combinations of N=14 subject groups; we calculated 15 average coupling functions and compared the similarity index of each in respect of the full coupling function, so as to calculate the variance of this 15 $\rho$ indexes.

Fig.\ \ref{Fig_7var} shows the variance dependence on the reduced number of subjects N. One can notice that the variance is relatively low. The dependence on N shows that the variance is low for reducing N until N=10 (perhaps N=9), after which for lower N the variance is rapidly increased. Therefore, the full number of sample size N=15 subjects is quite robust and has no big effect on the averaged coupling function. The results in Fig.\ \ref{Fig_7var} were calculated for two regions (delta 58 and alpha 21), but our investigation on other regions showed similar results for the variance where it was low for N=10 and then it got rapidly increased.     

}
\blue{\section{Association of region numbers to appropriate brain regions}}

Table~\ref{tab:table1} shows the relationship between the ordinal numbers of the regions as used in this paper and the designations of the regions according to the Desikan-Kiliany anatomical parcellation.
\begin{table*} [h!]
\caption{\label{tab:table1}Relationship between the ordinal numbers of the regions as used in this paper and the designations of the regions according to the Desikan-Kiliany anatomical parcellation.}
\begin{ruledtabular}
\begin{tabular}{cccc||cccc}
Region number & Brain region & Hemisphere & Brain lobe & Region number & Brain region & Hemisphere & Brain lobe\\
\hline
$	1	$	&	banksst	&	left	&	temporal	& $	35	$	&	banksst	&	right	&	temporal	\\
$	2	$	&	caudalanteriorcingulate	&	left	&	cingulate frontal & $	36	$	&	caudalanteriorcingulate	&	right	&	cingulate frontal\\
$	3	$	&	caudalmiddlefrontal	&	left	&	frontal	 & $	37	$	&	caudalmiddlefrontal	&	right	&	frontal	\\
$	4	$	&	cuneus	&	left	&	occipital	& $	38	$	&	cuneus	&	right	&	occipital	\\
$	5	$	&	entorhinal	&	left	&	temporal	& $	39	$	&	entorhinal	&	right	&	temporal	\\
$	6	$	&	fusiform	&	left	&	temporal	& $	40	$	&	fusiform	&	right	&	temporal	\\
$	7	$	&	inferiorparietal	&	left	&	parietal	& $	41	$	&	inferiorparietal	&	right	&	parietal	\\
$	8	$	&	inferiortemporal	&	left	&	temporal	& $	42	$	&	inferiortemporal	&	right	&	temporal	\\
$	9	$	&	isthmuscingulate	&	left	&	cingulate parietal	& $	43	$	&	isthmuscingulate	&	right	&	cingulate parietal	\\
$	10	$	&	lateraloccipital	&	left	&	occipital	&  $	44	$	&	lateraloccipital	&	right	&	occipital	\\
$	11	$	&	lateralorbitofrontal	&	left	&	frontal	 & $	45	$	&	lateralorbitofrontal	&	right	&	frontal	\\
$	12	$	&	lingual	&	left	&	occipital	&  $	46	$	&	lingual	&	right	&	occipital	\\
$	13	$	&	medialorbitofrontal	&	left	&	frontal	 & $	47	$	&	medialorbitofrontal	&	right	&	frontal	\\
$	14	$	&	middletemporal	&	left	&	temporal	& $	48	$	&	middletemporal	&	right	&	temporal	\\
$	15	$	&	parahippocampal	&	left	&	temporal	& $	49	$	&	parahippocampal	&	right	&	temporal	\\
$	16	$	&	paracentral	&	left	&	frontal	 & $	50	$	&	paracentral	&	right	&	frontal	\\
$	17	$	&	parsopercularis	&	left	&	frontal	 &  $	51	$	&	parsopercularis	&	right	&	frontal	\\
$	18	$	&	parsorbitalis	&	left	&	frontal	 &  $	52	$	&	parsorbitalis	&	right	&	frontal	\\
$	19	$	&	parstriangularis	&	left	&	frontal	 & $	53	$	&	parstriangularis	&	right	&	frontal	\\
$	20	$	&	pericalcarine	&	left	&	occipital	& $	54	$	&	pericalcarine	&	right	&	occipital	\\
$	21	$	&	postcentral	&	left	&	parietal	&  $	55	$	&	postcentral	&	right	&	parietal	\\
$	22	$	&	posteriorcingulate	&	left	&	cingulate parietal	& $	56	$	&	posteriorcingulate	&	right	&	cingulate parietal	\\
$	23	$	&	precentral	&	left	&	frontal	 & $	57	$	&	precentral	&	right	&	frontal	\\
$	24	$	&	precuneus	&	left	&	parietal	& $	58	$	&	precuneus	&	right	&	parietal	\\
$	25	$	&	rostralanteriorcingulate	&	left	&	cingulate frontal & $	59	$	&	rostralanteriorcingulate	&	right	&	cingulate frontal\\
$	26	$	&	rostralmiddlefrontal	&	left	&	frontal	 & $	60	$	&	rostralmiddlefrontal	&	right	&	frontal	\\
$	27	$	&	superiorfrontal	&	left	&	frontal	 &  $	61	$	&	superiorfrontal	&	right	&	frontal	\\
$	28	$	&	superiorparietal	&	left	&	parietal	& $	62	$	&	superiorparietal	&	right	&	parietal	\\
$	29	$	&	superiortemporal	&	left	&	temporal	& $	63	$	&	superiortemporal	&	right	&	temporal	\\
$	30	$	&	supramarginal	&	left	&	parietal	& $	64	$	&	supramarginal	&	right	&	parietal	\\
$	31	$	&	frontalpole	&	left	&	frontal	 & $	65	$	&	frontalpole	&	right	&	frontal	\\
$	32	$	&	temporalpole	&	left	&	temporal & $	66	$	&	temporalpole	&	right	&	temporal	\\
$	33	$	&	transversetemporal	&	left	&	temporal & $	67	$	&	transversetemporal	&	right	&	temporal	\\
$	34	$	&	insula	&	left	&	temporal	& $	68	$	&	insula	&	right	&	temporal	\\
\end{tabular}
\end{ruledtabular}
\end{table*}

\end{document}